\documentclass{elsart}

\usepackage{amsmath}
\usepackage{amsfonts}
\usepackage{graphicx}

\newcommand{\average}[1]{\langle #1\rangle}

\newcommand{\eq}[1]{(\ref{#1})}
\newcommand{\Eq}[1]{Eq.~\eq{#1}}

\begin{document}
\begin{frontmatter}
\title{Superstatistical generalization of the work fluctuation theorem}
\author[1,2]{C. Beck}
\author[3]{and E.G.D. Cohen}
\address[1]{Kavli Institute for Theoretical Physics,
University of California at Santa Barbara, Santa Barbara, CA
93106-4030, USA}
\address[2]{School of Mathematical Sciences, Queen Mary, University
of London, Mile End Road, London E1 4NS, UK}
\address[3]{
The Rockefeller University, 1230 York Avenue, New York, New York
10021, USA}

\date{15 December 2003}

\begin{abstract} We derive a generalized version of the work
fluctuation theorem for nonequilibrium systems with
spatio-temporal temperature fluctuations. For $\chi^2$-distributed
inverse temperature we obtain a generalized fluctuation theorem
based on $q$-exponentials, whereas for other temperature
distributions more complicated formulae arise. Since
$q$-exponentials have a power law decay, the decay rate in this
generalized fluctuation theorem is much slower than the
conventional exponential decay. This implies that work
fluctuations can be of relevance for the design of micro or nano
structures, since the work done on the system is relatively much
larger than in the conventional fluctuation theorem.
\end{abstract}
\end{frontmatter}

\section{Introduction}

In this paper two recent developments are being combined:
Fluctuation theorems \cite{ecomo} --- \cite{cili} and Tsallis
statistics (and further generalized statistics) in non-equilibrium
stationary states \cite{tsa1}
--- \cite{reynolds}.

The conventional fluctuation theorems give the ratio of the
probability $P(W_\tau)$ to find a fluctuation of, e.g., the work
$W_\tau$ done on a system over a time interval $\tau$ to the
probability $P(-W_\tau)$ of an equal amount of work $-W_\tau$ done
by the system. This ratio is given by an exponential function of
$W_\tau$, i.e.\ one has $P(W_\tau)/P(-W_\tau) = \exp(\beta
W_\tau)$, where $\beta$ is the inverse temperature. On the other
hand, the 'superstatistical' generalization, derived here, can
yield for this ratio (for example) a power law instead of
exponential behavior.

Superstatistics is a 'statistics of a statistics' relevant for
driven nonequilibrium systems with a stationary state \cite{super}
--- \cite{reynolds}. One assumes that the system has spatio-temporal temperature
fluctuations on a relatively large scale, due to external forces
acting on the system. One statistics is then given by ordinary
Boltzmann factors (in space-time cells where the temperature is
sufficiently constant), and the other one by the probability
distribution of $\beta$ in the various cells. It has been shown
that these types of models generate Tsallis statistics if $\beta$
is $\chi^2$-distributed \cite{wilk,prl}. But many other
generalized statistics are possible as well
\cite{sattin,super,cast}.

We will present a superstatistical generalization of the
fluctuation theorem. Our result could have important consequences
for the work fluctuations in micro or nano systems, because the
ratio $P(-W_\tau)/P(W_\tau)$ of the above mentioned probabilities
is much larger for the superstatistical case. In fact, this
implies that for large fluctuations a micro or nano structure will
be subjected relatively to much more work than in the conventional
case. This could, in particular, be of relevance if micro or nano
systems are placed into a fluctuating or turbulent environment,
with {\em externally} produced temperature fluctuations due to
constantly acting external driving forces. Moreover, besides the
work fluctuation theorem, there is also a heat fluctuation theorem
\cite{prl-zoco}. For small systems the heat fluctuations allowed
by the heat fluctuation theorem can also lead to {\em internal}
fluctuations of the temperature. All this illustrates the need to
consider fluctuation theorems for systems where the temperature is
not constant but fluctuates.

The Fluctuation Theorem can be visualized by considering a very
long trajectory in the phase space of the system in a
non-equilibrium stationary state, then partitioning this
trajectory in segments, on all of which the system spends an equal
time $\tau$. A histogram is then made of $W_\tau$ and $W_{-\tau}$
for all segments, which will allow an (approximate) determination
of the probabilities $P(W_\tau)$ and $P(-W_\tau)$ and their ratio.

The main difference between the conventional and the
superstatistical Fluctuation Theorem is that e.g. in the case of a
Brownian particle, as considered here, the friction and the noise
strength in the Langevin equation are constants, while in the case
of superstatistics these quantities can fluctuate in space and
time. A concrete physical interpretation of the Langevin equation
with constant coefficients in terms of electric circuits has been
proposed in \cite{cili}. In this approach, the velocity of the
Brownian particle corresponds to the time-dependent current
measured in the circuit. Applying this physical interpretation to
our Langevin equation with fluctuating coefficients, our
superstatistical fluctation theorem describes the work
fluctuations in a small circuit system whose surrounding
temperature fluctuates on a relatively large time scale.

In this paper we will only be concerned with fluctuations of the
work done on or by a system in a non-equilibrium stationary state
in the large time $t$ limit. The fluctuations of the heat are much
more complicated and will not be considered here \cite{prl-zoco}.
There are very many other similar fluctuation theorems for work
and heat e.g. for transient states and for finite times, all of
which could be generalized to superstatistical fluctuation
theorems. As said, we will restrict ourselves here to the simple
case, in order to elucidate the principle of the superstatistical
generalization of the conventional fluctuation theorems on the
above described model: the replacement of an exponential by e.g.\
a power law.

This paper is organized as follows. In section 2 we formulate a
superstatistical Langevin model for the joint process of velocity
and position of a Brownian particle moving in a changing
environment. In section 3 we consider the overdamped case of that
model, which leads to a superstatistical Langevin equation for the
position only. In section 4 we derive the superstatistical version
of the work fluctuation theorem. Finally, in section 5 some
concrete examples of superstatistical fluctuation theorems are
written down.

\section{Superstatistical Langevin model}

Consider a spherical Brownian particle in three dimensions with
radius $R$ and mass $m$. We assume that this moves in a fluid of
viscosity $\eta$ in an environment that has spatio-temporal
temperature fluctuations. These temperature fluctuations take
place on a rather large spatio-temporal scale and are produced by
external driving forces that act on the system and change the
environment of the Brownian particle. A typical example could be a
particle moving in a turbulent flow experiment. An effective
statistical mechanics description of the test particle is given by
a superstatistics \cite {super}, a 'statistics of a statistics',
where one is given by a probability density associated with the
Brownian particle for a locally constant temperature, while the
other one is given by the distribution $f(\beta)$ of the inverse
temperature $\beta$ in the various spatio-temporal regions (cells)
of approximately constant $\beta$. The temperature $\beta^{-1}$
surrounding the Brownian particle on its way through these spatial
regions of size $L^3$ varies on the rather large time scale
$\tau_L$ (due to both, the spatial movement of the particle and
also due to a possible explicit time dependendence of the
temperature field). $\tau_L$ is the typical time scale the
Brownian particle spends in the cell of size $L^3$. We assume that
the relaxation time $\tau_r$ of the Brownian particle satisfies
$\tau_r <<\tau_L$ so that local equilibrium can be established.

To establish a fluctuation theorem, we assume that the motion of
the Brownian particle is restricted by a time-dependent harmonic
potential \cite{Wang} . For $t\leq0$ the minimum $\mathbf x_t^*$
of the harmonic potential is at the origin, $\mathbf x^*_t=0$,
whereas for $t>0$ it moves with a velocity $\mathbf v^*_t$. This
velocity $\mathbf v^*_t$ can, in principle, be an arbitrary
function of time \cite{zon}. The equations of motion for the
particle are
\begin{subequations}
\begin{eqnarray}
    \dot{\mathbf x}_t &=& \mathbf v_t
\label{dotx}
\\
    m\dot{\mathbf v}_t &=& -\alpha\mathbf v_t
                -k(\mathbf x_t-\mathbf x^*_t)
                + \zeta_t,
\label{dotv}
\end{eqnarray}
\end{subequations}
where $\mathbf x_t$ and $\mathbf v_t$ are the position and velocity of
the Brownian particle, respectively.

There are
three different forces acting on the particle. Firstly, the damping force
$-\alpha\mathbf v_t$, with Stokes' law yielding $\alpha$ as
\begin{equation}
    \alpha=6\pi\eta R.
\label{Stokes}
\end{equation}
Secondly, there is the harmonic force $-k(\mathbf x_t -\mathbf x^*_t)$.
Finally, there is the
random force $\zeta_t$, which is taken
to be Gaussian white noise:
\begin{equation}
    \average{\zeta_t} = 0;
\quad
    \average{\zeta_t\zeta_s} = 2\beta^{-1}\alpha\delta(t-s)
\label{sf}
\end{equation}
The important point is that $\beta$ is not assumed to be constant,
but varies on the rather large time scale $\tau_L$.

The local equilibrium distribution, for some given local $\beta$,
assuming that this $\beta$ persists locally for a sufficiently
long time interval for the Brownian particle to reach local
equilibrium, is given by
\begin{equation}
   f_{leq, \beta}(\mathbf x,\mathbf v)
    = \left(\frac{\beta\sqrt{km}}{2\pi}\right)^{3}
            e^{
            -\beta\left(\frac12 m|\mathbf v|^2
            + \frac12 k|\mathbf x-\mathbf x_t^*|^2\right)
            } .
\label{equilibrium0}
\end{equation}
In the long-term, the particle moves through spatial regions which
have different inverse temperatures $\beta$. Denoting the
probability density of $\beta$ by $f(\beta)$, one asymptotically
obtains a convolution of the various equilibrium distributions for
different $\beta$. The observed long-term distribution (the
marginal distribution where $\beta$ is not fixed anymore), relevant
for $t>>\tau_L$, is
obtained by integrating over all $\beta$:
\begin{equation}
p(\mathbf x, \mathbf v)=\int_0^\infty f(\beta) f_{leq,
\beta}(\mathbf x,\mathbf v) d\beta.  \label{margi}
\end{equation}

Let us consider an important example, namely a
$\chi^2$-distributed $\beta$ \cite{hast,wilk,prl,super}. This
means the probability density of $\beta$ is given by
\begin{equation}
f (\beta) = \frac{1}{\Gamma \left( \frac{n}{2} \right)} \left\{
\frac{n}{2\beta_0}\right\}^{\frac{n}{2}} \beta^{\frac{n}{2}-1}
\exp\left\{-\frac{n\beta}{2\beta_0} \right\} \label{fluc} .
\end{equation}
$n$ denotes the number of independent Gaussians $X_i$ contributing
to the $\chi^2$-distribution of $\beta=\sum^n_{i=1} X^2_i$. The
mean of the inverse temperature is given by
\begin{equation}
\langle \beta \rangle =\int_0^\infty\beta f(\beta) d\beta= \beta_0
\end{equation}
and the variance by
\begin{equation}
\langle \beta^2 \rangle -\beta_0^2= \frac{2}{n} \beta_0^2.
\end{equation}
Let us define the energy of the Brownian particle as
\begin{equation}
E=\frac{1}{2}m|\mathbf v|^2+\frac{1}{2}k |\mathbf x -\mathbf x_t^*|^2.
\end{equation}
The integration in Eq.~(\ref{margi}) can be performed explicitly,
and one obtains \cite{prl} the generalized canonical distributions
of nonextensive statistical mechanics \cite{tsa1,tsa3,tsa2,abe}
\begin{equation}
p(\mathbf x, \mathbf v) \sim \frac{1}{\left(
1+\tilde{\beta}(q-1)E\right)^{\frac{1}{q-1}}} \label{10}
\end{equation}
with
the following identifications
\begin{equation}
q=1+\frac{2}{n+6} \label{previousqq}
\end{equation}
and
\begin{equation}
\tilde{\beta}=\frac{\beta_0}{1-3(q-1)}.
\end{equation}
The marginal distribution (\ref{10}) describing the long-term
distribution of position and momentum of the particle becomes a
so-called $q$-exponential \cite{tsa3,abe}. For other distributions
$f(\beta)$, one obtains other superstatistics, which all reduce to
Tsallis statistics for sharply peaked temperature distributions,
as shown in \cite{super}.

\section{Simplification for the overdamped case}
To proceed to a simpler stochastic model, we may proceed in the
usual way \cite{vanKampen}  and consider the system in the
overdamped case
\begin{equation}
    mk\ll \alpha^2.
\label{overdamped}
\end{equation}
In that limit one obtains a Langevin equation for the
position only,
\begin{equation}
    \dot{\mathbf x}_t = -\tau_r^{-1} (\mathbf x_t-\mathbf x^*_t)
                + \alpha^{-1}\zeta_t,
\label{langevin}
\end{equation}
with relaxation time
\begin{eqnarray}
\tau_r &=& \frac{\alpha}{k}.
\label{taurdef}
\end{eqnarray}
The local equilibrium distribution reduces to
\begin{equation}
    p_{eq, \beta}(\mathbf x)
=\int\! d\mathbf v\, f_{eq}(\mathbf x,\mathbf v)
    = (k\beta/2\pi)^{3/2}
            e^{-\beta\frac{k}{2}|\mathbf x-\mathbf x_t^*|^2},
\label{equilibrium}
\end{equation}
and again the marginal distribution
is given by
\begin{equation}
p(\mathbf x)= \int_0^\infty f(\beta)p_{eq,\beta}(\mathbf x) d\beta .
\end{equation}
For the example of a $\chi^2$-distributed inverse temperature,
we obtain

\begin{equation}
p(\mathbf x) \sim \frac{1}{\left( 1+\frac{1}{2}k\tilde{\beta}(q-1)
|\mathbf x -\mathbf x_t^*|^2\right)^{\frac{1}{q-1}}}
\end{equation}
with
\begin{equation}
q=1+\frac{2}{n+3} \label{previousq}
\end{equation}
and
\begin{equation}
\tilde{\beta}=\frac{\beta_0}{1-\frac{3}{2}(q-1)}.
\end{equation}

\section{Generalized work fluctuation theorem}

In this section we follow reference \cite{zon} with appropriate
modifications to solve our superstatistical model.

The work $W_\tau$ done on the system during a time interval $\tau$
is given by
\begin{equation}
    W_\tau = \int_0^\tau\!dt\, \mathbf v^*_t\cdot
        [-k(\mathbf x_t-\mathbf x^*_t)]. \label{22}
\end{equation}

The conventional fluctuation theorem for stationary states with a
given constant inverse temperature $\beta$ reads for sufficiently
large $\tau$:
\begin{equation}
        \frac{P(-W_\tau)}{P(W_\tau)} = e^{-\beta W_\tau}.
        \label{TFT1}
\end{equation}

In \cite{zon} the
fluctuation theorem was derived from a Langevin model with
constant temperature. We now extend this approach to
nonequilibrium systems with spatio-temporally fluctuating
temperatures.

Let us first restrict to spatial regions (cells) where the
temperature can be taken to be locally constant. Within these
cells, all relevant random variables are Gaussians provided $\tau
< \tau_L$. In Eq.~(\ref{22}), $W_\tau$ is a linear function of
$\mathbf x_t$. Combined with the Gaussian nature both of the
Green's function of the Ornstein-Uhlenbeck process and of the
initial distribution, this implies that the distribution $P$ of
$W_\tau$ in a given cell for a given time interval $\tau$ is
Gaussian as well:
\begin{equation}
    P(W_\tau) = \frac{1}{
    \sqrt{2\pi V_\tau}}
    \exp \left\{-\frac{(W_\tau-M_\tau)^2}{2V_\tau}\right\}
\label{PT}
\end{equation}
Here $M_\tau$ denotes the mean of $W_\tau$ and $V_\tau$ the
variance of $W_\tau$.

Following section C in \cite{zon}, we obtain straightforwardly for
sufficiently large $\tau$
\begin{equation}
    M_\tau = k \int_0^\tau\!dt_2'\int_0^{t_2'}\!dt'_1\,
    e^{-(t_2'-t_1')/\tau_r}
    \mathbf v^*_{t_2'}\cdot\mathbf v^*_{t_1'}
\label{MT}
\end{equation}
and
\begin{equation}
    V_\tau = 2 \beta^{-1} M_\tau.
\label{VTMT}
\end{equation}
This relation proves a {\em local} fluctuation theorem in the
cells with constant $\beta$. Combining \Eq{PT} and \Eq{VTMT}, we
may write
\begin{equation}
P(W_\tau)=\sqrt{\frac{\beta}{4\pi M_\tau}}\exp \left\{
-\frac{\beta}{4M_\tau}(W_\tau-M_\tau)^2 \right\}
\end{equation}
and
\begin{equation}
P(-W_\tau)=\sqrt{\frac{\beta}{4\pi M_\tau}}\exp \left\{
-\frac{\beta}{4M_\tau}(W_\tau + M_\tau)^2 \right\},
\end{equation}
which immediately leads to
\begin{equation}
        \frac{P(-W_\tau)}{P(W_\tau)} = e^{-\beta W_\tau}.
\label{TFT}
\end{equation}

Let us now take into account the temperature fluctuations. We have
the following ordering of time scales
\begin{equation}
\tau_r << \tau <\tau_L << t,
\end{equation}
where $\tau_r$ is the relaxation time to local equilibrium, $\tau$
is the time scale of a local fluctuation theorem, $\tau_L$ is the
time scale the particle spends in a cell of size $L^3$, and $t$ is
the total time elapsed. Two different experimental setups seem
possible, for which we derive two different versions of a
generalized fluctuation theorem for $t\to \infty$.

{\bf Setup 1.}  In the experiment one measures the ratio
$P(-W_\tau)/P(W_\tau)$ in local time intervals $\tau < \tau_L$,
within which the temperature stays sufficiently constant. In this
case one obtains in the long-term ($t>> \tau$)
\begin{equation}
\langle \frac{P(-W_\tau)}{P(W_\tau)} \rangle = \int_0^\infty
f(\beta) e^{-\beta W_\tau} d\beta . \label{setup11}
\end{equation}
Here the expectation $\langle \cdots \rangle$ denotes an
expectation with respect to the distribution $f(\beta)$.

{\bf Setup 2.} In a different experimental setup $P(W_\tau)$ and
$P(-W_\tau)$ are each measured over a very long time period
that includes many different states with different
temperatures. In this case one obtains in the long-term ($t>>\tau$)
\begin{equation}
\frac{\langle P(-W_\tau)\rangle }{\langle
P(W_\tau)\rangle}=\frac{\int_0^\infty f(\beta) \exp \left\{
-\frac{\beta}{4M_\tau}(W_\tau+M_\tau)^2 \right\} \sqrt{\beta}
d\beta}{\int_0^\infty f(\beta) \exp \left\{
-\frac{\beta}{4M_\tau}(W_\tau-M_\tau)^2\right\}
\sqrt{\beta}d\beta} .
\end{equation}

\section{Some examples}

\subsection{$\chi^2$-distribution}

The assumption of a $\chi^2$-distributed inverse temperature
$\beta$ of the form (\ref{fluc}) leads to a Tsallis-generalized
fluctuation theorem. One obtains for an experimental setup of type
1 a generalized fluctuation theorem
of the form
\begin{equation}
\langle \frac{P(-W_\tau)}{P(W_\tau)} \rangle =\int_0^\infty
e^{-\beta W_\tau} f(\beta ) d\beta =(1+\tilde{\beta} (q-1) W_\tau
)^{- \frac{1}{q-1}}, \label{tsallis-d}
\end{equation}
with
\begin{equation}
q=1+\frac{2}{n} \label{qf}
\end{equation}
and
\begin{equation}
\tilde{\beta}=\beta_0.
\end{equation}
This means the ordinary exponential $e^x$ in the fluctuation
theorem is replaced by a $q$-expontial $e_q^x:=(1-(q-1)x)^{
-\frac{1}{q-1}}$ \cite{tsa2,abe}. Note that in the fluctuation
theorem there is now an asymptotic power law in $W_\tau$ for large
$W_\tau$, which means that events with negative work-production,
i.e. with work done on the system, are now much more likely than
in the case without $\beta$ fluctuations. Thus rare events are
significantly enhanced by the temperature fluctuations: previously
they were exponentially suppressed, now they obey a power law.
This can be of potential interest for small electrical circuits
\cite{cili} in a fluctuating environment \cite{super}. Also note
that the $q$ relevant for the generalized fluctuation theorem as
given by eq.~(\ref{qf}) is different from the one for the
nonextensive equilibrium distribution given in
Eq.~(\ref{previousqq}) or Eq.~(\ref{previousq}), due to the fact
that for the superstatistical invariant distributions we integrate
over $\beta$-dependent normalization factors, whereas for the
fluctuation theorem (in setup 1) we do not.

For small $W_\tau$, the result (\ref{tsallis-d}) can also be
written as the following power expansion (cf.\cite{super},
eq.~(14)):
\begin{equation}
\langle \frac{P(-W_\tau)}{P(W_\tau)} \rangle = e^{-\beta_0 W_\tau}
(1+\frac{1}{2} (q-1) \beta_0^2 W_\tau^2 -
\frac{1}{3}(q-1)^2\beta_0^3W_\tau^3+...) \label{4}
\end{equation}

For setup 2 we obtain the formula
\begin{equation}
\frac{\langle P(-W_\tau)\rangle}{\langle P(W_\tau) \rangle} =
\left( \frac{1+\tilde{\beta}(q-1)\frac{(W_\tau
-M_\tau)^2}{4M_\tau}}{
(1+\tilde{\beta}(q-1)\frac{(W_\tau+M_\tau)^2}{4M_\tau}}\right)^\frac{1}{q-1}
\end{equation}
where
\begin{equation}
q=1+\frac{2}{n+1}
\end{equation}
and
\begin{equation}
\tilde{\beta}=\frac{\beta_0}{1-\frac{1}{2}(q-1)}.
\end{equation}

\subsection{Log-normal distribution}

The log-normal distribution
\begin{equation}
f(\beta) = \frac{1}{\beta s \sqrt{2\pi}}\exp\left\{ \frac{-(\log
\frac{\beta}{m})^2}{2s^2}\right\}
\end{equation}
yields yet another possible superstatistics. $m$ and $s$ are
parameters. The average $\beta_0$ of the above log-normal
distribution is given by $\beta_0=m\sqrt{w}$ and the variance by
$\sigma^2=m^2w(w-1)$, where $w:= e^{s^2}$ \cite{super}. Lognormal
superstatistics has applications in Lagrangian turbulence models
\cite{euro,reynolds}. Experimentally measured data of
single-particle accelerations are described very well by these
models. The physical meaning of the variables occurring in the
turbulence models is different, for example the variable $\mathbf
x$ in the Langevin equation is representing the acceleration of a
test particle in the turbulent flow, rather than the position of a
Brownian particle, while $\beta$ is not related to the inverse
physical temperature in the flow but to the fluctuating energy
dissipation in a cascade. Nevertheless, the mathematics is the
same and one obtains in experimental setup 1 a fluctuation theorem
of the form (\ref{setup11}), which for small $W_\tau$ has the
power law expansion (cf.\cite{super}, eq.(16))
\begin{equation}
\langle \frac{P(-W_\tau)}{P(W_\tau)}\rangle =e^{-\beta_0 W_\tau}
[1+\frac{1}{2}m^2w(w-1) W_\tau^2+ O(W_\tau^3)].
\end{equation}
One might speculate that these or similar types of fluctuation
theorems describe the backward scattering in the turbulent
energy cascade.

\subsection{F-distribution}

Consider a $\beta\in [0,\infty]$ distributed according to the
F-distribution \cite{hast,sattin,super}
\begin{equation}
f(\beta) =\frac{\Gamma ((v+w)/2)}{\Gamma (v/2) \Gamma (w/2)}
\left( \frac{bv}{w} \right)^{v/2} \frac{\beta^{\frac{v}{2}-1}}{(1+
\frac{vb}{w}\beta)^{(v+w)/2}} .
\end{equation}
Here $w$ and $v$ are positive integers and $b>0$ is a parameter.
The average of $\beta$ is given by
\begin{equation}
\beta_0=w/b(w-2)
\end{equation}
and the variance by
\begin{equation}
\sigma^2=2w^2(v+w-2)/b^2v(w-2)^2(w-4) .
\end{equation}
For small $W_\tau$ we obtain in experimental setup 1 an expansion
of the form (cf.\cite{super}, eq.~(20))
\begin{equation}
\langle \frac{P(-W_\tau)}{P(W_\tau)} \rangle =e^{-\beta_0W_\tau} [
1+\frac{1}{2} \sigma^2 W_\tau^2 + O(W_\tau^3)] .
\end{equation}

\subsection{Sharply peaked distributions}

While in general the large $W_\tau$ behavior in the various
fluctuation theorems depends strongly on the superstatistics
considered (i.e.\ on the function $f(\beta)$), the small-$W_\tau$
behavior is universal, in the sense that there is always the same
quadratic correction term to the ordinary fluctuation theorem. To
see this we can adopt a proof previously described in
\cite{super}. For any distribution $f(\beta)$ with average
$\beta_0 :=\langle \beta \rangle$ and variance $\sigma^2:=\langle
\beta^2 \rangle -\beta_0^2$ we can write
\begin{eqnarray}
\langle \frac{P(-W_\tau)}{P(W_\tau)} \rangle &=& \langle e^{-\beta
W_\tau} \rangle = e^{-\beta_0 W_\tau} e^{+\beta_0 W_\tau} \langle
e^{-\beta W_\tau}\rangle \nonumber
\\ &=& e^{-\beta_0 W_\tau} \langle e^{-(\beta -\beta_0)W_\tau}\rangle
\nonumber \\ &=& e^{-\beta_0 W_\tau} \left(1 +\frac{1}{2}\sigma^2
W_\tau^2 +\sum_{r=3}^\infty \frac{(-1)^r}{r!} \langle
(\beta-\beta_0)^r \rangle W_\tau^r \right).
\end{eqnarray}
The coefficients of the powers $W_\tau^r$ are the $r$-th moments
of the distribution $f(\beta)$ about the mean $\beta_0$. For small
enough $\sigma W_\tau$, in setup 1 the first order correction term
to the ordinary fluctuation term is always quadratic in $W_\tau$:
\begin{equation}
\langle \frac{P(-W_\tau)}{P(W_\tau)} \rangle \approx e^{-\beta_0
W_\tau } (1+\frac{1}{2}\sigma^2 W_\tau^2)
\end{equation}

For sharply peaked distributions in $\beta$, the fluctuation
theorem for experimental setup 2 also simplifies and
one obtains
\begin{equation}
\frac{\langle P(-W_\tau) \rangle}{\langle P(W_\tau) \rangle}
\approx e^{-\beta_0 W_\tau} \left( \frac{1+\frac{\sigma^2}{32}
\frac{(W_\tau+M_\tau)^4}{M_\tau^2}}{ 1+\frac{\sigma^2}{32}
\frac{(W_\tau-M_\tau)^4}{M_\tau^2}} \right) \approx e^{-\beta_0
W_\tau}\left(1+\frac{\sigma^2}{4} \frac{W_\tau}{M_\tau}
(W_\tau^2+M_\tau^2) \right) ,
\end{equation}
where $\beta_0$ and $\sigma^2$ are the mean and variance of the
probability density $\tilde{f}(\beta):=C\beta^{1/2}f(\beta)$ used
for mapping type-B superstatistics into type-A superstatistics
\cite{super}.

\newpage

{\underline{Acknowledgement}}

C. B. acknowledges support in part by the National Science
Foundation under Grant No. PHY99-07949 and E. G. D. C. of the
Office of Basic Energy Science of the U.S. Department of Energy
under grant number DE-FG02-88-13847.  E. G. D. C. also
acknowledges C. Tsallis' constant questions about the possibility
of a power law Fluctuation Theorem and the very helpful assistance
of R. van Zon.


\begin{thebibliography}{10}



\bibitem{ecomo}
D.~J. Evans, E.~G.~D. Cohen, and G.~P. Morriss, Phys. Rev. Lett. {\bf 71},
  2401  (1993).


\bibitem{Evansetal94}
D.~J. Evans and D.~J. Searles, Phys. Rev. {E} {\bf 50},  1645  (1994).


\bibitem{GallavottiCohen95a}
G. Gallavotti and E.~G.~D. Cohen, Phys. Rev. Lett. {\bf 74},  2694  (1995).


\bibitem{GallavottiCohen95b}
G. Gallavotti and E.~G.~D. Cohen, J. Stat. Phys. {\bf 80},  931  (1995).


\bibitem{SearlesEvans00}
D.~J. Searles and D.~J. Evans, J. Chem. Phys. {\bf 113},  3503  (2000).


\bibitem{Kurchan98}
J. Kurchan, J. Phys. {A}, Math. Gen. {\bf 31},  3719  (1998).


\bibitem{LebowitzSpohn99}
J.~L. Lebowitz and H. Spohn, J. Stat. Phys. {\bf 95},  333  (1999).


\bibitem{otherexperiment}
S. Ciliberto and C. Laroche, J. Phys. IV, France {\bf 8},  215  (1998).


\bibitem{Wang}
G.~M. Wang {\it et~al.}, Phys. Rev. Lett. {\bf 89},  050601  (2002).

\bibitem{CohenGallavotti99}
E.~G.~D. Cohen and G. Gallavotti, J. Stat. Phys. {\bf 96},  1343
(1999).

\bibitem{zon} R. van Zon and E.G.D. Cohen,
Phys. Rev. {\bf 67E}, 046102 (2003)

\bibitem{prl-zoco} R. van Zon and E.G.D. Cohen,
Phys. Rev. Lett. {\bf 91}, 110601 (2003)

\bibitem{cili} R. van Zon, S. Ciliberto, and E.G.D. Cohen,
cond-mat/0311629


















\bibitem{tsa1} C. Tsallis, J. Stat. Phys. {\bf 52}, 479 (1988)
\bibitem{tsa3} C. Tsallis, Braz. J. Phys. {\bf 29}, 1 (1999)
\bibitem{wilk} G. Wilk and Z. Wlodarczyk, Phys. Rev. Lett. {\bf 84},
2770 (2000)
\bibitem{prl} C. Beck, Phys. Rev. Lett. {\bf 87}, 180601 (2001)
\bibitem{cohen} E.G.D. Cohen, Physica {\bf 305A}, 19 (2002)
\bibitem{tsa2} C. Tsallis,  R.S. Mendes and A.R. Plastino, Physica {\bf 261A},
534 (1998)
\bibitem{abe} S. Abe, Y. Okamoto (eds.), Nonextensive
Statistical Mechanics and Its Applications, Springer, Berlin
(2001)
\bibitem{BLS} C. Beck, G.S. Lewis, H.L. Swinney, Phys. Rev. E {\bf 63},
035303(R) (2001)
\bibitem{sattin} F. Sattin and L. Salasnich,
Phys. Rev. {\bf 65E}, 035106(R) (2002)
\bibitem{super} C. Beck and E.G.D. Cohen,
Physica {\bf 322A}, 267 (2003)
\bibitem{souza} C. Tsallis and A.M.C. Souza, Phys. Rev. {\bf 67E},
026106 (2003)
\bibitem{euro} C. Beck, Europhys. Lett. {\bf 64}, 151 (2003)
\bibitem{reynolds} A. Reynolds, Phys. Rev. Lett. {\bf 91}, 084503 (2003)

\bibitem{cast} B. Castaing, Y. Gagne, E.J. Hopfinger,
Physica D {\bf 46}, 177 (1990)


\bibitem{hast} N.A.J. Hastings and J.B. Peacock,
{\em Statistical Distributions}, Butterworth, London (1974)
\bibitem{vanKampen}
N.~G. van Kampen, {\em Stochastic Processes in Physics and
Chemistry}, revised
  and enlarged ed. (North Holland, Amsterdam, 1992)

\end{thebibliography}
\end{document}